\documentclass[11pt]{article}
\usepackage{amssymb,amsmath}

\newcommand{\be}{\begin{equation}}
\newcommand{\ee}{\end{equation}}
\newcommand{\bea}{\begin{eqnarray}}
\newcommand{\eea}{\end{eqnarray}}
\def\1#1{^{(#1)}}
\def\la{\langle}
\def\ra{\rangle}
\def\be{\begin{equation}}
\def\ee{\end{equation}}
\def\bea{\begin{eqnarray}}
\def\eea{\end{eqnarray}}

\def\DEV[#1]{\mathaccent'27{#1}}
\begin{document}
\title{Dispersion and Dispersivity Tensors in \\
Saturated Porous Media with Uniaxial Symmetry}
\author{Leonid G. Fel and Jacob Bear\\
\\Department of Civil and Enviromental Engineering, \\
Technion -- Israel Institute of Technology, Haifa 3200, Israel} \vspace{-.2cm}
\date{}
\maketitle
\begin{abstract}
The coefficients of dispersion,  $D_{ij}$, and the dispersivity, $a_{ijkl}$, appear
in the expression for the flux of a solute in saturated flow through porous media.
We present a detailed analysis of these tensors in an axially symmetric porous
medium and show that in such a medium, the dispersivity is governed by 6 independent
moduli. We present also the constraints that have to be satisfied by these moduli.
We also show that at least two independent experiments are required in order to
obtain the values of these coefficients for any three-dimensional porous medium
domain.
\end{abstract}
\section{Introduction}\label{a1}

The coefficients of dispersion and dispersivity appear in the expression for the
flux, $\bf J$, of a solute in saturated flow through porous media. Here, a porous
medium domain is one that is made up of two parts: a solid matrix and a void-space
completely occupied by one or more fluid phases. The void-space is interconnected,
thus allowing fluid flow through it \cite{be72}. Both parts are distributed over the
entire porous medium domain. We study the transport of a solute dissolved in a fluid
phase. At every point within the fluid, a solute concentration, $\gamma$, and a
fluid velocity, $\bf V$, can be identified. A mathematical model describing the
solute's transport within the void space requires information on the fluid-solid
interface that bounds the fluid occupied domain. However, in natural materials, such
as soil and fractured rock, this information is not available. Therefore, the
problem of solute transport is usually described in terms of averaged velocity,
$\overline {\bf V}$, and averaged solute concentration, $\overline\gamma$. These
averaged values, taken over some representative elementary volume, are assigned to
every point in the porous medium domain.

With the total (advective) solute flux at a point inside the void-space defined
by $\gamma{\bf V}$, and at a point in the porous medium domain, regarded as a
continuum, defined by $\overline\gamma\overline{\bf V}$, we have \cite{be72},
\[
\overline{\gamma{\bf V}} = {\overline\gamma}{\overline{\bf V}}+\overline{\gamma'
{\bf V}'},\quad\mbox{where}\quad \gamma'=\gamma-\overline{\gamma},\quad
{\bf V}'={\bf V}-\overline{{\bf V}},
\]
are deviations from the average values, with $\overline{\gamma'}=\overline{{\bf
V}'}=0$. The above equality states that the total flux of the solute at a point in
the porous medium domain, is made up of an advective flux (= product of average
concentration and average velocity), and another flux, called dispersive flux, equal
to the average of the product of the two deviations, ${\bf J}= \overline{\gamma'{\bf
V}'}$.

The above discussion can be extended to any extensive quantity, with $\gamma$
representing its intensive quantity. Thus, the phenomenon of dispersion occurs
whenever an extensive quantity  (mass of a solute, in the example considered above)
is transported in a fluid phase that flows in the void space of a porous medium. It
is a consequence of using an average velocity and average concentration to describe
the advective flux of the solute in the void space. The total flux is the sum of the
advective flux and the dispersive one. In this article, we use the  mass of
dissolved chemical species as the considered extensive quantity. For short, we
denote the average values: $\overline{\gamma}=c$, $\overline{{\bf V}}={\bf U}$.

The coefficient of dispersion, $D_{ij}$, is a symmetric tensor of 2nd rank, which
appears in the Fickian-type expression for the dispersive flux of a solute (mass of
solute per unit area of void space in a cross-section of the porous medium)
\cite{be72},
  \bea
  J_i=-\sum_{j=1}^3D_{ij}\nabla_j c,\quad D_{ij}=D_{ji},\quad i,j=1,2,3\,
  \label{q1}
  \eea
where $J_i$ denotes the $i$th component of the solute's 3-dim flux vector,
${\bf J}$, and $c$ is the solute's average concentration. Equation (\ref{q1})
is valid for the general case of any anisotropic porous medium, with isotropic
media as a special case.

In contrast to similar linear flux laws, e.g., Ohm's and Fourier's laws, with
corresponding tensorial coefficients of electroconductivity, $\sigma_{ij}$, and of
thermoconductivity, $\lambda_{ij }$, respectively, which depend only on properties
of the considered medium, here, in the case of the dispersive flux in saturated flow
through a porous medium, the dispersion tensor $D_{ij}$ depends also on the fluid's
average velocity ${\bf U}$ in the void space \cite{be61},
 \bea
 D_{ij}=\frac1{U}\sum_{k,l=1}^3a_{ijkl}\;U_kU_l,\quad\;\;\;{\bf U}^2=\sum_{k=1}
^3 U_k^2,\quad\;\;\;{\bf U}^2=U^2,\label{q2} \eea
where $a_{ijkl}$ is a property, called dispersivity, of the porous medium only, and
$U_i$ is the $i$th component of the fluid's average velocity vector $\bf U$. The
dispersivity, is a 4th rank tensor, with intrinsic symmetry $\left[V^2\right]^2$ in
Jahn notations \cite{ja49}
   \bea
\left[V^2\right]^2\;:\;\;a_{ijkl}=a_{ijlk}\;,\;\;\;\;a_{ijkl}=a_{jikl}.
   \label{q3}
    \eea

Additional simplifications in dealing with the dispersivity tensor can be obtained
by making use of certain auxiliary algebraic relations among its components. These
relations appear under the action of the crystallographic point symmetry group
${\cal G}\subset {\cal O}(3)$ of the porous medium that makes some (not all) of the
36 elements of $a_{ijkl}$ survive, or at least independent of the remaining ones.
According to \cite{ss84}, the decrease in the number $N({\cal G})$ of
$a_{ijkl}$-components is significant when we increase the symmetry, i.e.,
  \bea
N({\cal C}_1)=36\;\rightarrow\;\ldots\;\rightarrow\;N({\cal D}_{2h})=12\;
\rightarrow\;N({\cal D}_{4h})=7\;\rightarrow\;N({\cal D}_{\infty h})=6\;
\rightarrow\;N({\cal O}(3))=2.\label{q3a}
    \eea

In thermodynamics, the rate of entropy production, $\sigma$, is related to the
thermodynamic driving force, ${\bf X}$, and to the thermodynamic flux, ${\bf
Y}$. According to \cite{gm62}, two vectors ${\bf X}$ and ${\bf Y}$ are
considered conjugated if they obey the relationship
  \bea
    \sigma=\sum_{i=1}^3Y_iX_i.\label{q4}
      \eea
In the case considered here, the dispersive flux of the solute, ${\bf J}$, is driven
by ${\bf \nabla} c$, which acts as a `driving force.'  In other words, ${\bf X}=g
\,{\bf \nabla} c$ and ${\bf Y}=g \, {\bf J}$, where $g$ denotes a (dimensional)
scalar parameter, which depends on the considered transport phenomenon. In this
case, the rate of entropy production is expressed by
 \bea
  \sigma=g^2\sum_{i,j=1}^3D_{ij}\;\nabla_i c\;\nabla_j c.\label{q5}
     \eea

By requiring that the quadratic form (\ref{q5}) be positive definite, we derive
the following constraints on the $D_{ij}$-matrix,
 \bea
  D_{ii}> 0,\;\;\det\left\{\!\begin{array}{ll}\!D_{11}&D_{12}\\
  \!D_{21}&D_{22}\end{array}\!\right\}> 0,\;\;
   \det\left\{\!\begin{array}{ll}D_{11}&D_{13}\\ D_{31}&D_{33}\end{array}\!
    \right\}> 0,\;\;\det\left\{\!\begin{array}{ll}D_{22}& D_{23}
     \\D_{23}&D_{33}\end{array}\!\right\}> 0,\;\;
      \det D_{ij}> 0.\nonumber
 \eea
The above constraints hold irrespective of the fluid's averaged velocity ${\bf
U}$. Furthermore, by (\ref{q2}), they provide another set of constraints
imposed on elements of the dispersivity tensor, $a_{ijkl}$. Such straightforward
way to establish these constraints is not effective, since it requires a lot of
routine algebra and does not prevent repetitive inequalities. In section
\ref{a4}, we shall present a much more effective method for deriving the sought
constraints.

In the meantime, let us estimate the total number of such inequalities. Making
use of the 2-dim matrix representation of the $a_{ijkl}$-tensor \cite{ss84},
one can show that the number $M({\cal G})$ of such constraints decreases also
with growth of the symmetry ${\cal G}$ of the porous medium,
 \bea
M({\cal C}_1)=63\;\rightarrow\;\ldots\;\rightarrow\;M({\cal D}_{2h})=10\;
\rightarrow\;M({\cal D}_{4h})=7\;\rightarrow\;M({\cal D}_{\infty h})=6\;
\rightarrow\;M({\cal O}(3))=2.\label{q3b}
   \eea
\section{The Tensor $a_{ijkl}$ in a Porous Medium with Uniaxial Symmetry}
\label{a2}

In this section, we study the 3-dim tensor $a_{ijkl}$ whose components depend only
on porous medium properties, actually, for saturated flow, only on the geometry of
the medium's void space. We shall focus on porous media with uniaxial symmetry. Our
interest in this kind of porous medium is motivated by two reasons. First, in
saturated flow in ground water geological formations, called aquifers, the porous
medium comprising the latter is often axially symmetric \cite{ma86}, with a vertical
axis, e.g., in a layered aquifer. On the other hand, four decades of continuing
discussions (see \cite{li02} and references therein) on explicit expressions of the
tensors $a_{ijkl}$ and $D_{ ij}$ in porous media with uniaxial symmetry have not
provided any rigorous analysis of these entities.

Instead of a 2-dim matrix representation of the $a_{ijkl}$-tensor, presented in
\cite{ss84}, we suggest another representation, which originates in the theory of
uniaxial nematic liquid crystals (LC). The underlying idea is that when we compare
the $a_{ijkl}$-tensor with the 4th rank viscosity tensor, $\eta_{ijkl}$, of a
uniaxial LC, its intrinsic symmetry is $\left[\left[V^2\right]^2\right]$, i.e.,
slightly `stronger' than (\ref{q3}), with,
  \bea
\left[\left[V^2\right]^2\right]\;:\;\quad\eta_{ijkl}=\eta_{ijlk},\quad
\eta_{ijkl}=\eta_{jikl},\quad \eta_{ijkl}=\eta_{klij}.\label{q6}
     \eea

According to \cite{l86}, the known representation of $\eta_{ijkl}$ is
  \bea
\eta_{ijkl}&=&\eta_1\delta_{ij}\delta_{kl}+\eta_2\left(\delta_{ik}\delta_{jl}+
\delta_{il}\delta_{jk}\right)+\eta_{34}\left(e_ie_j\delta_{kl}+e_ke_l\delta_{ij}
\right)\nonumber\\
&& + \eta_5\left(e_ie_k\delta_{jl}+e_je_k\delta_{il}+e_ie_l\delta_{jk}+e_je_l
\delta_{ik}\right)+\eta_6e_ie_je_ke_l,  \label{q7}
    \eea
in which the $e_i$'s denote the three components of the unit vector ${\bf e}$,
along the symmetry axis of the medium.

From (\ref{q7}), it follows that there exists only one term, $(e_ie_j\delta_{
kl}+e_k e_l\delta_{ij})$, which  reduces its intrinsic symmetry: $\left[\left[
V^2\right]^2\right]\to\left[V^2\right]^2$, after decomposition of this term into
two separate terms: $e_ie_j\delta_{kl}$ and $e_ke_l\delta_{ij}$. This explains
why, in accordance with (\ref{q3b}), the tensor  $a_{ijkl}$ has six independent
moduli and how to construct its most generic form:
 \bea
a_{ijkl}&=&a_1\;\delta_{ij}\delta_{kl}+\frac{a_2}{2}\left(\delta_{ik}\delta_{jl}
+\delta_{il}\delta_{jk}\right)+a_3\;e_ie_j\delta_{kl}+a_4\;e_ke_l\delta_{ij}+
\nonumber\\
&&\frac{a_5}{2}\left(e_ie_k\delta_{jl}+e_je_k\delta_{il}+e_ie_l\delta_{jk}+
  e_je_l\delta_{ik}\right)+a_6\;e_ie_je_ke_l.\label{q8}
    \eea
Note that by making $e_i\to 0$ in (\ref{q8}), we obtain the dispersivity tensor
for an isotropic porous medium,
   \bea
a_{ijkl}=a_1\delta_{ij}\delta_{kl}+\frac{a_2}{2}\left(\delta_{ik}\delta_{jl}+
\delta_{il}\delta_{jk}\right).\label{q9}
    \eea

The axisymmetric tensor (\ref{q8}) has to be supplemented by a set of
constraints that results from the thermodynamic requirement imposed on the
rate of entropy production, $\sigma$, defined in  (\ref{q5}).

\section{The Tensor $D_{ij}$ in a Porous Medium with Uniaxial Symmetry}
\label{a3}

In this section, we consider the 3-dim tensor $D_{ij}$ and certain interesting
related phenomena. Based on (\ref{q2}) and (\ref{q8}), we rewrite this tensor as
  \bea
   UD_{ij}=a_1{\bf U}^2\delta_{ij}+a_2U_iU_j+a_3{\bf U}^2e_ie_j+a_4\la {\bf e},
{\bf U}\ra^2\delta_{ij}+a_5\la {\bf e},{\bf U}\ra(U_ie_j+e_iU_j)+a_6\la{\bf e},
{\bf U}\ra^2 e_ie_j,\label{q10}
     \eea
where $\la{\bf e},{\bf U}\ra$ denotes a scalar product of the two vectors:
${\bf e}$ and ${\bf U}$.

It is interesting to note that Batchelor \cite{ba46}, in his work on
axisymmetric turbulence, derived a general expression for $D_{ij}$ that is
based on five tensorial terms: $e_ie_j$, $\delta_{ij}$, $U_iU_j$, $U_ie_j$,
$U_je_i$, and four scalar functions, $C_i({\bf U},{\bf e})$,
 \bea
   UD_{ij}=C_1({\bf U},{\bf e})e_ie_j+C_2({\bf U},{\bf e})\delta_{ij}+
C_3({\bf U}, {\bf e})U_iU_j+C_4({\bf U},{\bf e})(U_ie_j+U_je_i).\label{q11}
    \eea

Keeping in mind that there exist only two quadratic-in-$U$ scalar invariants,
$U^2$ and $\la{\bf U},{\bf e}\ra^2$, and only one linear-in-$U$ scalar
invariant, $\la{\bf U},{\bf e}\ra$, we can verify the equivalence of the two
expressions, (\ref{q10}) and (\ref{q11}), by noting that
 \bea
  C_1({\bf U},{\bf e})=a_3U^2+a_6\la {\bf U},{\bf e}\ra^2,\;C_2({\bf U},
{\bf e})=a_1U^2+a_4\la {\bf U},{\bf e}\ra^2,\;C_3({\bf U},{\bf e})=a_2,\;
C_4({\bf U}, {\bf e})=a_5\la {\bf U},{\bf e}\ra.\nonumber
 \eea

We introduce the two 6-dim vectors, ${\bf D}_6$ and ${\bf a}_6$,
 \bea
{\bf D}_6=\{D_{11},D_{22},D_{33},D_{12},D_{23},D_{31}\},\quad {\bf a}_6=\{
a_1,a_2,a_3,a_4,a_5,a_6\},\label{q12}
     \eea
and denote their transposed counterpartners by ${\bf D}_6^T$ and ${\bf a}_6^T$.
Next, we rewrite (\ref{q10}) in the form
  \bea
  {\bf D}_6^T=\frac1{U}\;\Delta_1\left({\bf U},{\bf e}\right)\cdot {\bf a}_6^T,
  \label{q13}
     \eea
where the $6\times 6$ matrix $\Delta_1\left({\bf U},{\bf e}\right)$ is defined as
\bea \Delta_1\left({\bf U},{\bf e}\right)=\left\{\begin{array}{cccccc} {\bf U}^2&
U_1^2&e_1^2{\bf U}^2&\la {\bf e},{\bf U}\ra^2&2\la {\bf e},{\bf U}\ra e_1U_1&
\la {\bf e},{\bf U}\ra^2e_1^2\\
{\bf U}^2&U_2^2&e_2^2{\bf U}^2&\la {\bf e},{\bf U}\ra^2&2\la {\bf e},{\bf U}
\ra e_2U_2&\la {\bf e},{\bf U}\ra^2e_2^2\\
{\bf U}^2&U_3^2&e_3^2{\bf U}^2&\la {\bf e},{\bf U}\ra^2&2\la {\bf e},{\bf U}
\ra e_3U_3&\la {\bf e},{\bf U}\ra^2e_3^2\\
0&U_1U_2&e_1e_2{\bf U}^2&0&\la{\bf e},{\bf U}\ra(U_1e_2+e_1U_2)&\la {\bf e},
{\bf U}\ra^2e_1e_2\\
0&U_2U_3&e_2e_3{\bf U}^2&0&\la{\bf e},{\bf U}\ra(U_2e_3+e_2U_3)&\la {\bf e},
{\bf U}\ra^2e_2e_3\\
0&U_3U_1&e_3e_1{\bf U}^2&0&\la{\bf e},{\bf U}\ra(U_3e_1+e_3U_1)&\la {\bf e},
{\bf U}\ra^2e_3e_1\end{array}\right\}.
   \label{q14}
      \eea

Equation (\ref{q13}) opens the way to the derivation of the 6 dispersivity
$a_i$-moduli from values of $D_{ij}$. The latter can be obtained from
experimental measurements, assuming that the symmetry axis ${\bf e}$ is known,
and so is the velocity ${\bf U}$, in an appropriate reference frame, $x_1,x_2,
x_3$. However, a straightforward calculation shows that the matrix $\Delta_1
\left({\bf U},{\bf e}\right)$ is degenerated,
 \bea
  \det \Delta_1\left({\bf U},{\bf e}\right)=0,\quad \mbox{rank}\;\Delta_1
    \left({\bf U},{\bf e}\right)=4\;. \label{q15}
     \eea
Equality (\ref{q15}) leads to two most important consequences and raises a
question:
\begin{enumerate}
\item At most, four independent components $D_{ij}$ can be obtained from a
 single experiment. The remaining two $D_{ij}$'s are linearly representable by
 the first four.
\item One cannot get all six dispersivity $a_i$-moduli  from a single
  experiment.
\item How many experiments do we have to perform in order to obtain all six
$a_i$-values?
\end{enumerate}
In what follows, we shall demonstrate that two different experiments suffice for
obtaining all six  $a_i$-values. We consider two cases (= experiments):
\begin{enumerate}
\item $U_1=0$, $U_2=U_3=u$ and $e_1=1$, $e_2=e_3=0$. Then,
\bea
D_{11}=2(a_1+a_3)u,\quad D_{22}=D_{33}=(2a_1+a_2)u, \quad D_{23}=a_2u,\quad
  D_{12}=D_{31}=0.\label{b14a}
 \eea
 The above relationships provide three moduli: $a_1$, $a_2$ and
$a_3$.
\item $U_1=U_2=u$, $U_3=0$ and $e_1=1$, $e_2=e_3=0$. Then,
\bea
&&D_{11}=(2a_1+a_2+2a_3+a_4+2a_5+a_6)u,\quad D_{12}=(a_2+a_5)u,\label{b14}\\
&&D_{22}=(2a_1+a_2+a_4)u,\quad D_{33}=(2a_1+a_4)u,\quad D_{23}=D_{31}=0\;.
\nonumber
 \eea
\end{enumerate}
The above relations provide three additional moduli: $a_4$, $a_5$ and $a_6$.

Thus, by two experiments, with different setups of ${\bf U}$, and a known
${\bf e}$, we have obtained the entire set of  6 dispersivity moduli in a 3-dim
saturated porous medium with uniaxial symmetry.

\subsection{The Tensor $D_{ij}$ in an Isotropic Porous Medium}\label{a31}

For an isotropic porous medium, instead of (\ref{q10}), we have
 \bea
   UD_{ij}=a_1{U}^2\delta_{ij}+a_2U_iU_j.\label{q16}
       \eea
This means that only two moduli: $a_1$ and $a_2$ are required for a full 
description of the $D_{ij}$-components. The expression for $\Delta_1\left({\bf 
U},{\bf e}\right)$-matrix, which is defined in (\ref{q14}), is even more 
degenerated, with $rank\Delta_1\left({\bf U}, {\bf e}\right)=2$. This indicates 
a strong linear dependence among the $D_{ij}$-values: there exist four linear 
relations among the six $D_{ij}$-elements.

Let us consider an isotropic porous medium, with the setup: $U_1=u$, $U_2=U_3
=0$. Then,
 \bea
((D_{ij}))=u\left\{\begin{array}{ccc}a_1+a_2&0&0\\0&a_1&0\\0&0&a_1\end{array}
\right\}.\label{q17}
    \eea

In ground water hydrology \cite{be72}, the modulus $a_1$ is called transversal
dispersivity (denoted  $a_T$) and $a_1+a_2$ is called longitudinal
dispersivity (denoted $a_L$).

\section{Thermodynamic constraints}\label{a4}

In this section we determine the algebraic constraints that have to be satisfied by
the six dispersivity $a_i$-moduli for an axisymmetric porous medium. We substitute
(\ref{q10}) into (\ref{q5}) and present $\sigma$ in its invariant form, irrespective
of the reference frame,
 \bea
\frac{\sigma}{g^2}&=&a_1\;{\bf U}^2 \left({\bf \nabla} c\right)^2+a_2\;
\la{\bf U},{\bf \nabla} c\ra^2+a_3\;{\bf U}^2 \la{\bf e},{\bf \nabla} c\ra^2+
a_4\;\la{\bf e},{\bf U}\ra^2\left({\bf \nabla} c\right)^2\label{q18}\\
&&\hspace{2cm} +2\;a_5\;\la{\bf e},{\bf U}\ra\la{\bf U},{\bf \nabla} c\ra\la
{\bf \nabla} c,{\bf e}\ra+a_6\;\la{\bf e},{\bf U}\ra^2\la{\bf e},{\bf\nabla}
c\ra^2\;.\nonumber
    \eea

We then represent (\ref{q18}) as a quadratic form. This is a standard way to write
the necessary and sufficient conditions for $\sigma$ to be positive definite. For
this purpose, we choose the reference frame such that $e_1=e_2=0$ and $e_3=1$. By
substituting these into (\ref{q18}), we get
 \bea
\frac{\sigma}{g^2}&=&a_1\sum_{i=1}^3U_i^2\sum_{i=1}^3\left(\nabla_i c\right)^2
+a_2\left(\sum_{i=1}^3U_i\nabla_i c\right)^2+a_3\left(\nabla_3 c\right)^2
\sum_{i=1}^3U_i^2+a_4U_3^2\sum_{i=1}^3\left(\nabla_i c\right)^2\nonumber\\
&&\hspace{2cm}+2\;a_5U_3\nabla_3 c\sum_{i=1}^3U_i\nabla_i c+a_6U_3^2 \left( \nabla_3
c\right)^2\;.\label{q19} \eea Expression (\ref{q19}) is a quadratic form in a 9-dim
space, with orthogonal basis $U_i\nabla_j c$, $i,j=1,2,3$. We decompose it in the
form
 \bea
\frac{\sigma}{g^2}&=&a_1\left[U_1^2\left(\nabla_2 c\right)^2+U_2^2\left(\nabla_1
c\right)^2\right]+(a_1+a_3)\left(\nabla_3 c\right)^2\left[U_1^2+U_2^2\right]+
(a_1+a_4)U_3^2\left[\left(\nabla_1 c\right)^2+\left(\nabla_2 c\right)^2
\right]\nonumber\\
&+&(a_1+a_2)\left[U_1^2\left(\nabla_1 c\right)^2+U_2^2\left(\nabla_2 c\right)
^2\right]+(a_1+a_2+a_3+a_4+2a_5+a_6)U_3^2\left(\nabla_3 c\right)^2\nonumber\\
&+&2a_2U_1U_2\nabla_1 c\nabla_2 c+2(a_2+a_5)U_3\nabla_3 c\left( U_1\nabla_1
c+U_2\nabla_2 c\right)\;,\label{q19a}
 \eea
and require the positive definiteness of the last expression. Then we obtain
three inequalities
\bea
  a_1> 0\;,\;\;\;a_1+a_3> 0\;,\;\;\;a_1+a_4> 0\;,\label{q19b}
   \eea
and requirement of positive definiteness of the $3\times 3$ matrix $\Gamma$, \bea
\Gamma=\left\{\begin{array}{cccccccc}
a_1+a_2&a_2&a_2+a_5\\a_2&a_1+a_2&a_2+a_5\\a_2+a_5&a_2+a_5&
a_1+a_2+a_3+a_4+2a_5+a_6\end{array}\right\}.\label{q19c} \eea The last claim leads
to three additional algebraically independent restrictions, \bea
&&a_1+a_2> 0\;,\;\;\;a_1+a_2+a_3+a_4+2a_5+a_6> 0\;,\label{q19d}\\
&&a_1^2+a_1(3a_2+a_3+a_4+2a_5+a_6)+2a_2(a_3+a_4+a_6)> 2a_5^2\;.\nonumber
\eea

Thus, we have arrived at 6 constraints imposed on the 6 dispersivity
$a_i$-moduli; this is in full agreement with (\ref{q3b}) for uniaxial symmetry
group ${\cal G}={\cal D}_{\infty h}$. Note that the first two inequalities in
(\ref{q19b}) and (\ref{q19d}) correspond to isotropic porous media.

\section{The Tensor $D_{ij}$ in a Planar Porous Medium with\\
Dihedral Symmetry}\label{a5}

In this section we consider the dispersion phenomenon in a 2-dim porous medium
domain, which is obtained by intersecting a 3-dim uniaxial porous medium with a
plane that is parallel to the uniaxial axis. In the resulting 2-dim porous medium
domain, there are two mirror lines, which are perpendicular to each other. By these
symmetry elements, it is easy to recognize the dihedral point symmetry group ${\cal
D}_2$, which is one of ten 2-dim point symmetry subgroups of ${\cal O}(2)$.

From the mathematical standpoint, the dispersion tensor, $D_{ij}$, can be
represented by a $2\times 2$ symmetric matrix, with three components: ${\bf
D}_3=\{D_{11},D_{22}, D_{12}\}$, instead of the $6\times 6$ matrix $\Delta_2
\left({\bf U},{\bf e}\right)$, as in (\ref{q13}). By doing so, we obtain:
 \bea
  {\bf D}_3^T=\frac1{U}\;\Delta_2\left({\bf U},{\bf e}\right)\cdot {\bf a}_6^T,
   \label{q20}
    \eea
where the $3\times 6$ matrix, $\Delta_2\left({\bf U},{\bf e}\right)$, is defined as
\bea \Delta_2\left({\bf U},{\bf e}\right)=\left\{\begin{array}{cccccc} {\bf
U}^2&U_1^2&e_1^2{\bf U}^2&\la {\bf e},{\bf U}\ra^2&2\la {\bf e},{\bf U}
\ra e_1U_1&\la {\bf e},{\bf U}\ra^2e_1^2\\
{\bf U}^2&U_2^2&e_2^2{\bf U}^2&\la {\bf e},{\bf U}\ra^2&2\la {\bf e},{\bf U}
\ra e_2U_2&\la {\bf e},{\bf U}\ra^2e_2^2\\
0&U_1U_2&e_1e_2{\bf U}^2&0&\la{\bf e},{\bf U}\ra(U_1e_2+e_1U_2)&\la{\bf e},
{\bf U}\ra^2e_1e_2\end{array}\right\}.\label{q21}
\eea

Equation (\ref{q20}) does not enable the determination of the six dispersivity
$a_i$-moduli, from 3 values of $D_{ij}$,  which can be obtained from a single
experiment. For comparison, in section \ref{a3}, we have shown that by two
experiments, with different setups of ${\bf U}$, and a known ${\bf e}$, we have
obtained the entire set of 6 dispersivity moduli in a 3-dim saturated porous medium
with uniaxial symmetry.

In this conjunction, the following question arises: how many experiments do we have
to perform in a 2-dim saturated porous medium domain with dihedral symmetry, in
order to obtain the entire set of 6 dispersivity moduli?

It is easy to show that two experiments are not enough. Indeed, let us consider
two different flow setups, with fluid averaged velocities ${\bf W}$ and ${\bf
U}$, and the same orientation of ${\bf e}$,
 \bea
_1{\bf D}_3^T=\frac1{U}\;\Delta_2\left({\bf W},{\bf e}\right)\cdot {\bf a}_6^T
   \;,\;\;\;\;\;\;_2{\bf D}_3^T=\frac1{U}\;\Delta_2\left({\bf U},{\bf e}\right)
\cdot {\bf a}_6^T.\label{q22}
   \eea
In (\ref{q22}) the front subscripts 1 and 2 in the notations $_1{\bf D}_3^T$ and
$_2{\bf D}_3^T$, indicate data obtained from the 1st and 2nd experiments, with
velocities ${\bf W}$ and ${\bf U}$, respectively. We now introduce a new 6-dim
vector, ${\sf D}_{6,a}$,
 \bea
{\sf D}_{6,a}=\{_1D_{11},\;_1D_{22},\;_1D_{12},\;_2D_{11},\;_2D_{22},
\;_2D_{12}\},\label{q23}
   \eea
composed of two 3-dim vectors $_1{\bf D}_3=\{_1D_{11},\;_1D_{22},\;_1D_{12}\}$ and
$_2{\bf D}_3=\{_2D_{11},\;_2D_{22}, \;_2D_{12}\}$, and rewrite the two equations in
(\ref{q22}), in the form
  \bea
{\sf D}_{6,a}^T=\frac1{U}\;\Delta_3\left({\bf W},{\bf U},{\bf e}\right)\cdot
  {\bf a}_6^T\;,\;\;\;\mbox{where}\label{q24}
    \eea
\bea
\Delta_3\left({\bf W},{\bf U},{\bf e}\right)=\left\{\begin{array}{cccccc}
{\bf W}^2&W_1^2&e_1^2{\bf W}^2&\la {\bf e},{\bf W}
\ra^2&2\la {\bf e},{\bf W}\ra e_1W_1&\la {\bf e},{\bf W}\ra^2e_1^2\\
{\bf W}^2&W_2^2&e_2^2{\bf W}^2&\la{\bf e},{\bf W}\ra^2&2\la {\bf e},{\bf
W}\ra e_2W_2&\la {\bf e},{\bf W}\ra^2 e_2^2\\
0&W_1W_2&e_1e_2{\bf W}^2&0&\la{\bf e},{\bf W}\ra(W_1e_2+e_1W_2)&\la{\bf e},
{\bf W}\ra^2e_1e_2\\
{\bf U}^2&U_1^2&e_1^2{\bf U}^2&\la {\bf e},{\bf U}\ra^2&2\la {\bf e},{\bf U}\ra
e_1U_1&\la {\bf e},{\bf U}\ra^2e_1^2\\{\bf U}^2&U_2^2&e_2^2{\bf U}^2&\la {\bf
e},{\bf U}\ra^2&2\la {\bf e},{\bf U}\ra e_2U_2&\la {\bf e},{\bf U}\ra^2e_2^2\\
0&U_1U_2&e_1e_2{\bf U}^2&0&\la{\bf e},{\bf U}\ra(U_1e_2+e_1U_2)&\la {\bf e},
{\bf U}\ra^2e_1e_2\end{array}\right\}.\nonumber
 \eea
A straightforward calculation shows that
 \bea
  \det\Delta_3\left({\bf W},{\bf U},{\bf e}\right)=0,\quad\mbox{rank}\;
   \Delta_3\left({\bf W},{\bf U},{\bf e}\right)=5.\label{q25}
    \eea

Thus, among the six elements of the vector ${\sf D}_{6,a}$, given in
(\ref{q23}), there are only five independent ones, and equation (\ref{q24})
does not suffice for providing the six dispersivity $a_i$-moduli.

We shall now show that three experiments are sufficient for providing the entire set
of $a_i$'s. Consider 3 different flow setups, with velocities: ${\bf W}$, ${\bf
U}$ and ${\bf \Upsilon}$ and with the same orientation of ${\bf e}$,
  \bea _1{\bf
D}_3^T=\frac1{U}\;\Delta_2\left({\bf W},{\bf e}\right)\cdot {\bf a}_6^T
\;,\;\;\;\;\; _2{\bf D}_3^T=\frac1{U}\;\Delta_2\left({\bf U},{\bf e}\right) \cdot
{\bf a}_6^T, \;,\;\;\;\;\; _3{\bf D}_3^T=\frac1{U}\;\Delta_2\left({\bf
\Upsilon},{\bf e}\right)\cdot {\bf a}_6^T,\label{q26}
   \eea
where $_1{\bf D}_3$ and $_2{\bf D}_3$ were defined earlir and $_3{\bf D}_3=\{
_3D_{11},\;_3D_{22}, \;_3D_{12}\}$. In contrast to (\ref{q23}), we introduce another
6-dim vector, ${\sf D}_{6,b}$,
 \bea {\sf
D}_{6,b}=\{_1D_{11},\;_1D_{12},\;_2D_{11},\;_2D_{12},\;_3D_{11},
  \;_3D_{12}\},\label{q27}
   \eea
composed of two elements, $_iD_{11}$ and $_iD_{12}$, of three 3-dim vectors
$_i{\bf D}_3$, $i=1,2,3$. We rewrite the three equations (\ref{q26}) in the
form,
  \bea
{\sf D}_{6,b}^T=\frac1{U}\;\Delta_4\left({\bf W},{\bf U},{\bf \Upsilon},
{\bf e}\right)\cdot {\bf a}_6^T\;,\;\;\;\mbox{where}\label{q28}
  \eea
  \bea
\Delta_4\left({\bf W},{\bf U},{\bf \Upsilon},{\bf e}\right)=
\left\{\begin{array}{cccccc} {\bf W}^2&W_1^2&e_1^2{\bf W}^2&\la {\bf e},{\bf
W}\ra^2&2\la {\bf e},{\bf W}\ra e_1W_1&\la {\bf e},{\bf W}\ra^2e_1^2\\
0&W_1W_2&e_1e_2{\bf W}^2&0&\la{\bf e},{\bf W}\ra(W_1e_2+e_1W_2)&\la{\bf e},
{\bf W}\ra^2e_1e_2\\
{\bf U}^2&U_1^2&e_1^2{\bf U}^2&\la {\bf e},{\bf U}\ra^2&2\la {\bf e},{\bf U}
\ra e_1U_1&\la {\bf e},{\bf U}\ra^2e_1^2\\
0&U_1U_2&e_1e_2{\bf U}^2&0&\la{\bf e},{\bf U}\ra(U_1e_2+e_1U_2)&\la {\bf e},
{\bf U}\ra^2e_1e_2\\
{\bf \Upsilon}^2&\Upsilon_1^2&e_1^2{\bf \Upsilon}^2&\la {\bf e},{\bf \Upsilon}
\ra^2&2\la {\bf e},{\bf \Upsilon}\ra e_1\Upsilon_1&\la {\bf e},{\bf \Upsilon}
\ra^2e_1^2\\
0&\Upsilon_1\Upsilon_2&e_1e_2{\bf \Upsilon}^2&0&\la{\bf e},{\bf \Upsilon}\ra(
\Upsilon_1e_2+e_1\Upsilon_2)&\la {\bf e},{\bf \Upsilon}\ra^2e_1e_2\end{array}
\right\}.\nonumber
 \eea

Note that the three 2-dim vectors: ${\bf W}$, ${\bf U}$ and ${\bf \Upsilon}$
are always coplanar. However, if at least two of them would be collinear, then
(\ref{q28}) cannot be solved uniquely, i.e. $\det\Delta_4\left(k{\bf U},{\bf
U},{\bf\Upsilon},{\bf e}\right)=0$, where $k$ is a real number. Otherwise, a
straightforward calculation shows that if neither two of the three vectors:
${\bf W}$, ${\bf U}$, and ${\bf \Upsilon}$ are collinear, then
\bea
\det\Delta_4\left({\bf W},{\bf U},{\bf \Upsilon},{\bf e}\right)\neq 0,\quad
\mbox{rank}\;\Delta_4\left({\bf W},{\bf U},{\bf \Upsilon},{\bf e}\right)=6.
\label{q29}
\eea
Thus, three special (not any) experiments suffice for providing all
six dispersivity $a_i$-moduli in a 2-dim saturated porous medium with dihedral
symmetry.

\section{Concluding Remarks}\label{a6}

For the case of an axisymmetric porous medium, we have shown that six
independent $a_i$-moduli are needed in order to determine all components of the
dispersivity tensor, and the constraints that these moduli have to satisfy. This
information is required when determining the latter by experiments for specific
porous media. We have also found the number of experiments that are required to
determine the entire set of dispersivity moduli for 2-dim and 3-dim domains.
This is important for experimental determination of the dispersivity.

\section{Acknowledgement}\label{a7}
The research was partly supported by the Kamea Fellowship program.

\vspace{2cm}

\date{\today}

e-mail: lfel@tx.technion.ac.il,

e-mail: cvrbear@tx.technion.ac.il
\end{document}